\documentclass[12pt,doublespace]{article}
\usepackage{graphicx}
\usepackage{epsfig}
\usepackage{float}
\usepackage{amssymb}
\usepackage[sort&compress,square]{natbib}
\usepackage{array}
\begin{document}
\newcolumntype{C}{>{\centering\arraybackslash}p{2cm}}

\title{Multifractal dynamics of stock markets}
\author{\L ukasz Czarnecki, Dariusz Grech\footnote{dgrech@ift.uni.wroc.pl} \ \\
Institute of Theoretical Physics\\
University of Wroc{\l}aw, PL-50-204 Wroc{\l}aw, Poland}

\date{}
\maketitle

\begin{abstract}
We present a comparative analysis of multifractal properties of financial time series built on stock indices from developing (WIG) and developed (S\&P500) financial markets. It is shown how the multifractal image of the market is altered with the change of the length of time series and with the economic situation on the market. We emphasize that the proper adjustment of scaling range for multiscaling power laws is essential to obtain the multifractal image of time series.
We analyze in this paper multifractal properties of real financial time series using H\"{o}lder $f(\alpha)$ representation and multifractal-DFA method. It is also investigated how multifractal properties of stocks change with variety of "surgeries" done on the initial real financial time series. This way we reveal main phenomena on the market influencing its multifractal dynamics. In particular, we focus on examining how multifractal picture of real time series changes when one cuts off extreme events like crashes or rupture points, and how fluctuations around the main trend in time series influence the multifractal behavior of financial series in the long-time horizon for both developed and developing markets.
\end{abstract}
$$
$$
\textbf{Keywords}: multifractality, econophysics, time series analysis, Hurst exponent\\
\textbf{PACS:}  89.65.Gh, 89.75.Da, 89.20.-a

\vspace{1cm}

\section{Introduction}
Multifractality [1-4] is rather a new concept when applied to financial time series. It may be considered as the higher order extension of the monofractal analysis used successfully e.g. in analysis of persistency level in data -- particularly in econophysics. So far it is not very clear what practical applications of multifractality in financial time series might be. However, some preliminary directions in this topic have already been shown (see e.g. [5]). In monofractal approach, one usually looks for the probabilistic (stochastic) behavior of some geometrical or topological properties of signal fluctuations around the local trend in given time series. One expects power-law relation between the quantity describing such fluctuations (e.g. variance) and the length of time window $s$ along which the fluctuation is being measured. A good example of such relation is the one provided in Detrended Fluctuation Analysis (DFA) [6-11] where the power law relation between the variance of detrended signal $F^2(s)$ and the width of time window $s$ reads
\begin{equation}
F^2(s) \sim s^{2H}
\end{equation}
with the Hurst scaling exponent $H$ ($0<H<1$) [12,13]. If $0<H<1/2$ the signal is said to be antipersistent, if $1/2<H<1$ the signal is persistent, while the case $H=1/2$ corresponds to no memory present in signal increments (Brownian motion).\\
The relation given by Eq.(1) is usually understood as independence of scaling properties of a system from the scale -- thus $H$ is constant. However, in many cases it might not be so. There might be a finite or infinite number of crossover points $s_X$ such that for time scales $s<s_X$ the fractal properties differ from these for $s>s_X$. In this case, the fractal structure will be described by the whole set of $H$ exponents instead of just one, thus revealing the multiple scaling rules. Using the standard DFA procedure one calculates only the leading scaling rule linked to major or more frequent fluctuations.
Moreover, distortion of the signal from the leading pattern before or below the crossover point $s_X$ may be very weak, so in order to extract them, one is forced to use more sophisticated technique based on the artificial amplification of weak (with respect to the average ones) fluctuations. Such philosophy makes the background of the modified DFA proposed by Kantelhardt et. al. and called multifractal DFA (MF-DFA) [14]. One replaces in this method the "ordinary" fluctuation function $F^2(s)$ from Eq.(1) by its q-th moment $F^2_q(s)$ defined as follows
\begin{equation}
F(s,q)=\{1/{2N}\sum^{2N}_k [F^2(s,k)]^{q/2}\}^{1/q}
\end{equation}
for $q\neq 0$, and
\begin{equation}
F(s,0)=\exp\{1/{4N}\sum^{2N}_k \ln[F^2(s,k)]\}
\end{equation}
for $q=0$, where $N$ counts the number of non-overlapping boxes of size $s$ each for which detrendization procedure is performed.\\
One obtains in this manner the whole continuous set of Hurst exponents $h(q)$ labeled by different $q\epsilon \bf{R}$. They are related to different level of amplification of small fluctuation in data. The $h(q)$ dependence is decreasing monotonic function of $q$ for stationary signal what basically reflects the fact that relatively small fluctuations happen more often in this signal than relatively big ones do [14]. If so, the so called H\"{o}lder spectrum or singularity spectrum $f(\alpha)$ [4] of the H\"{o}lder exponent $\alpha$ ca be used where
\begin{equation}
\alpha = \frac{ds(q)}{dq}
\end{equation}
and
\begin{equation}
s(q)=qh(q)-1
\end{equation}
while $f(\alpha)$ is determined by Legendre transform [16,17]
\begin{equation}
f(\alpha)=q(\alpha)\alpha-s(q(\alpha))
\end{equation}
The latter approach makes a useful representation of multifractality. The singularity spectrum in this case should have rather regular form of inverse parabolic shape as shown in Fig.2. The width of $f(\alpha)$ spectrum measures the multifractality level of the signal. For the pure monofractal signal this width converges to one point $\alpha=H=h(2)$ and $f(\alpha)=1$, while if the bigger amount of multifractality is present in the signal, the width of $f(\alpha)$ spectrum becomes wider.\\
We will analyze in this paper multifractal properties of real financial time series using $f(\alpha)$ representation and MF-DFA method argued as working better than other approaches [15]. Our main task is to investigate how multifractal properties of stocks change with variety of "surgeries" done on the initial real financial time series. We shall do this in following chapters. Our aim is to see what phenomena on the market influence its multifractal dynamics in the first place. In particular, we will focus on examining how multifractal picture of real time series is changed when one cuts off extreme events like crashes or rupture points, and how established trends or their absence influence the multifractal behavior of developed and developing markets.

\section{Multifractal noise in monofractal signal of finite length}

All statistical methods determining the fractal properties of time series give an exact result only for series of infinite length. Since this applies also to the MF-DFA scheme, one should know how the length of series is an important factor in discovering its multifractal structure. In fact, finite signals assumed by their construction to be monofractals, reveal some artificial multifractal structure. This is because big fluctuations in finite time series, measured within MF-DFA, seem to be more rare than for longer or infinite time series. It leads to smaller $h(q)$ value for $q\rightarrow\infty$ in finite series than for infinite ones, and finally, to some artificial multifractal structure of finite time series. This kind of multifractal noise should be subtracted first as a background influencing any real property of monofractal or multifractal time series of finite length.\\
To explore this issue we must have a ''test series`` with known fractal characteristic. We will use monofractal series generated by the random midpoint displacement (RMD) algorithm [22].
The resulting h(q) functions calculated within MF-DFA for three series of length $L=2^{14}$ generated with input Hurst exponent values $H=0.3, 0.5, 0.7$ respectively are presented in Fig. 1.\\
One may notice a small degree of multifractality (short span of $h(q)$) for all three series. The series with higher $H$ show a bigger deviation from monofractality.
The non-monotonic behaviour of $h(q)$ is reflected by the ''twist`` on top of the singularity spectrum (see Fig. 2). This effect will be also observed in following chapters where real financial data are analyzed.\\
One can also see that the maxima positions of the spectra reflect the assumed Hurst exponents values.
\\\indent It is essential to find the influence of series length on the width of multifractal spectrum. To observe this, one has to calculate the highest and lowest value of $\alpha$, as well as the position of the $f(\alpha)$ peak. This was done on a sample of 20 series each for four lengths $L=2^8, 2^{10}, 2^{12}, 2^{14}$ and for different $H$ values. The width of $f(\alpha)$ spectrum was rapidly changing with $L$ and seemed to be weakly dependent on $H$. Table 1 presents results found for $H=0.5$.
\begin{table}[ht]
\begin{center}
\begin{tabular}{|C|C|C|}
\hline
Series length $L$ & Spectrum width $\Delta\alpha$ & Peak position\\
\hline
$2^{8}$ & 0.73 & 0.56\\
$2^{10}$ & 0.28 & 0.50\\
$2^{12}$ & 0.21 & 0.50\\
$2^{14}$ & 0.22 & 0.51\\
\hline
\end{tabular}
\end{center}
\caption{An example of dependence between the length $L$ and the singularity spectrum width $\Delta\alpha$ for artificially generated RMD series with assumed input value of Hurst exponent $H=0.5$.}
\label{tab:rmd}
\end{table}
\\They confirm the significant narrowing of $\Delta\alpha$ spectrum width when the length $L$ of time series is increasing. Since we will deal later on with series lengths around $L\sim 2^{14}$, we may assume that the finite size effects should be around $\delta_{finite}(\Delta\alpha)\sim 0.20$.
\\\indent Before analyzing real data we also need to observe the scaling range of the $F(s,q) \sim s^{h(q)}$ relation. For the artificial data, this range spans almost the entire available time window lengths (see Fig. 3). However, for the entire set of real data (later sections will also discuss parts of these data), the scaling range becomes shorter and decreases with increasing  $\left|q\right|$ values. \\
The corresponding effect for the S\&P500 is visible in Fig. 4. It is even more significant for positive and larger $q$ what is shown in Fig. 5 (compare with Fig. 4). To solve this problem a careful separate study of all $F(s,q) \sim s^{h(q)}$ relations and manual selection of the scaling range is needed. It was done in analysis of all financial data presented below.

\section{Long-term data analysis}

The first and most obvious approach is to make the long-term data analysis, i.e., to calculate the multifractal structure of the stock market indices in the whole available time horizon. Our analysis focuses on two indices:
\begin{itemize}
\item WIG (Warsaw Stock Exchange)) - from April 16$^{th}$ 1991 till October 10$^{th}$ 2008
\item S\&P500 (NYSE \& Nasdaq) - from December 30$^{th}$ 1927 till September 3$^{rd}$ 2008
\end{itemize}
They are examples of developing (Poland) and developed (USA) markets respectively.\\
All time series were created from daily closing values of each index. The S\&P 500 data were downloaded from the financial web-site of Yahoo.com [18] and WIG data from a stock market web-site of Wirtualna Polska [19].
\\\indent The singularity spectra $f(\alpha)$ and $h(q)$ plots of all discussed time series (as well as their shuffles) are presented in Figs 6, 7. Surprisingly, all markets show unexpected non-monotonic  $h(q)$ behavior. This phenomenon, in addition to the problems with short scaling ranges (mentioned in the previous section), strongly suggests a disturbance in the multifractal structure of the series. Spectra of all three shuffled data sets are positioned around the expected location $\alpha = 0.5$. These sets of data also show a monotonic behavior of the $h(q)$ function, contrary to originally ordered data. The spectra width for original as well for shuffled data are much wider than those of the corresponding RMD spectra and hence indicate the multifractal content in dynamics of long-term market data. Simultaneously, the non-monotonic $h(q)$ behavior seems to indicate non-stationary effects in these data. The latter case needs more detailed study partly provided below.


\section {Multifractal analysis of financial data in established trends}

It has been shown by a number of authors (see e.g.[5,20,21]) that positive and negative fluctuations in time series have different fractal properties. O\'{s}wi\c{e}cimka et. al. has shown in Ref.[5] that DAX data in increasing and decreasing trends has different multifractal properties for particular choice of periods Dec.1997-Dec.1999 and May 2004-May 2006. It is quite natural to investigate whether these properties are general and if so, in what extent they are general for other markets. In particular, it is interesting  to know how multifractal properties of established trends depend on the internal structure of these trends. We were interested in the beginning  in positive (negative) long-lasting trends with no sub-trends in opposite direction. We tried to exactly repeat this way the analysis made for DAX in Ref.[5]. Such trends turned out to be very rare for S\&P 500 and WIG indices. The selected fragments of both indices with required properties are shown in Fig. 8. They have both long-lasting positive trend (bullish phase) followed by a long negative trend (bearish phase) and no astonishing internal events.\\
The resulting singularity spectra (see Fig.9) show the expected shift of the bearish phase spectra towards higher values of $\alpha$. Since these results were obtained from data with a much wider scaling range, they should be considered as much more reliable then previous ones presented for entire indices. It is also worth noting that the spectra are now much wider, which suggest a more complex multifractal structure. In the case of WIG this effect might be amplified by the short data set, nevertheless the spectra width of a RMD series with similar length is still narrower.\\
This improvement of singularity spectra for well determined trends in both indices (i.e. quasi quadratic dependence $f(\alpha)$ from $\alpha$) with respect to spectra obtained for the entire indices might suggest that the multifractality is strongly affected by non-stationarity of financial time series. To verify such hypothesis we checked what multifractal properties are hidden in subparts of entire index. The S\&P500 index is particularly good for this search because it is long enough comparing with WIG, thus containing much more amount of data. If one divides the entire S\&P500 history into two regimes - one with small fluctuations (small volatility) and the second one with high fluctuations (high volatility) as shown in Fig. 10, a very intriguing result for singularity spectra is found in both regimes. Results of this search are plotted in Figs.11, 12, where a comparison of singularity spectra and $h(q)$ dependence for both regimes with $f(\alpha)$ and $h(q)$ behavior for the entire index is provided. It is clear that both regimes  show separately a monotonic $h(q)$ dependence leading to reversed parabolic shape of $f(\alpha)$ . It is not the case of the entire index.\\
However, both spectra for two regimes are slightly different -- the width of multifractal spectrum (and the width of $h(q)$ dependence) is larger for the first regime characterized by lower volatility and no opposite sub-trends inside. We checked the same happened for other, even shorter trends, whenever they are affected by well formed sub-trends of opposite direction or by extreme events like crashes, etc. This phenomena is shown on example of S\&P500 time series without and with internal structure (see Figs.13, 14). Fig.13 presents an example of increasing trend with no visible distortion inside. We called it "good" data. Its dynamics is associated with nice multifractal properties - well established quadratic $f(\alpha)$ dependence and the monotonic $h(q)$ behaviour. Simultaneously, time series of the same length, but collecting data from the very next period, has some extreme events inside and it contains opposite sub-trends connected with these events (see Fig.14). We call them "bad" data. The scaling $F_q(s)\sim s^{h(q)}$ is worse here and the presence of "twist" in singularity spectrum $f(\alpha)$ confirms the non-monotonic character of $h(q)$ function and leads also to non-quadratic $f(\alpha)$ dependence. The multifractal character of time series is here very much affected by the presence of abrupt events, i.e. crashes or data with respectively large fluctuations.

\section{Influence of extreme events on multifractal dynamics of stocks}
The sub-trends containing abrupt events can  be artificially removed to see if such removal would improve the multiscaling properties of the new trend constructed after such "surgery". The applied procedure is simple. First, the time series with abrupt events is differentiated. Then, at the level of returns, these events are removed and the time series of remaining returns is integrated to get the new artificial evolution of financial index. Examples of such modification are visible for variety of events for both S\&P 500 and WIG indices in Figs.15, 16 and 17 respectively. It is remarkable, how much such surgery significantly improves multi-scaling properties of financial time series and restores the monotonic character of decreasing $h(q)$ function. The width of singularity spectrum has also noticeably increased when abrupt events had been removed.

\section{Conclusions}

We presented a comparative analysis of multifractal properties of financial time series built from stock indices of developing (WIG) and emergent (S\&P500) markets using MF-DFA technique. We found that financial time series in a very long time horizon show in both cases very short multiscaling ranges and in turn, their multifractal image is somehow obscured. Therefore we analyzed in shorter time horizon the selected parts of financial data chosen due to their specific properties.
\\\indent The most important result is that division of time series into regimes of distinctly different (large or small) fluctuations around the main trend, as well as removal of abrupt events like crashes or significant opposite  sub-trends, radically improves the multiscaling ranges  and restores the monotonic behavior of $h(q)$ dependence as well as the reversed parabolic shape of multifractal singularity spectrum $f(\alpha)$. The width of singularity spectrum also increases noticeably in these cases. This leads us to conclusion that non-stationarity of financial time series significantly influences their multifractal character leading to behavior beyond our expectation for multifractal spectrum shape known from studies with artificially created mono- and multifractal signals.
\\\indent It is clear that further research of length effects as well as the effects of non-stationarity in real time series influencing multiscaling and multifractality is needed with hope for further practical applications. \\

\begin{figure}[h]
\includegraphics[width=\columnwidth]{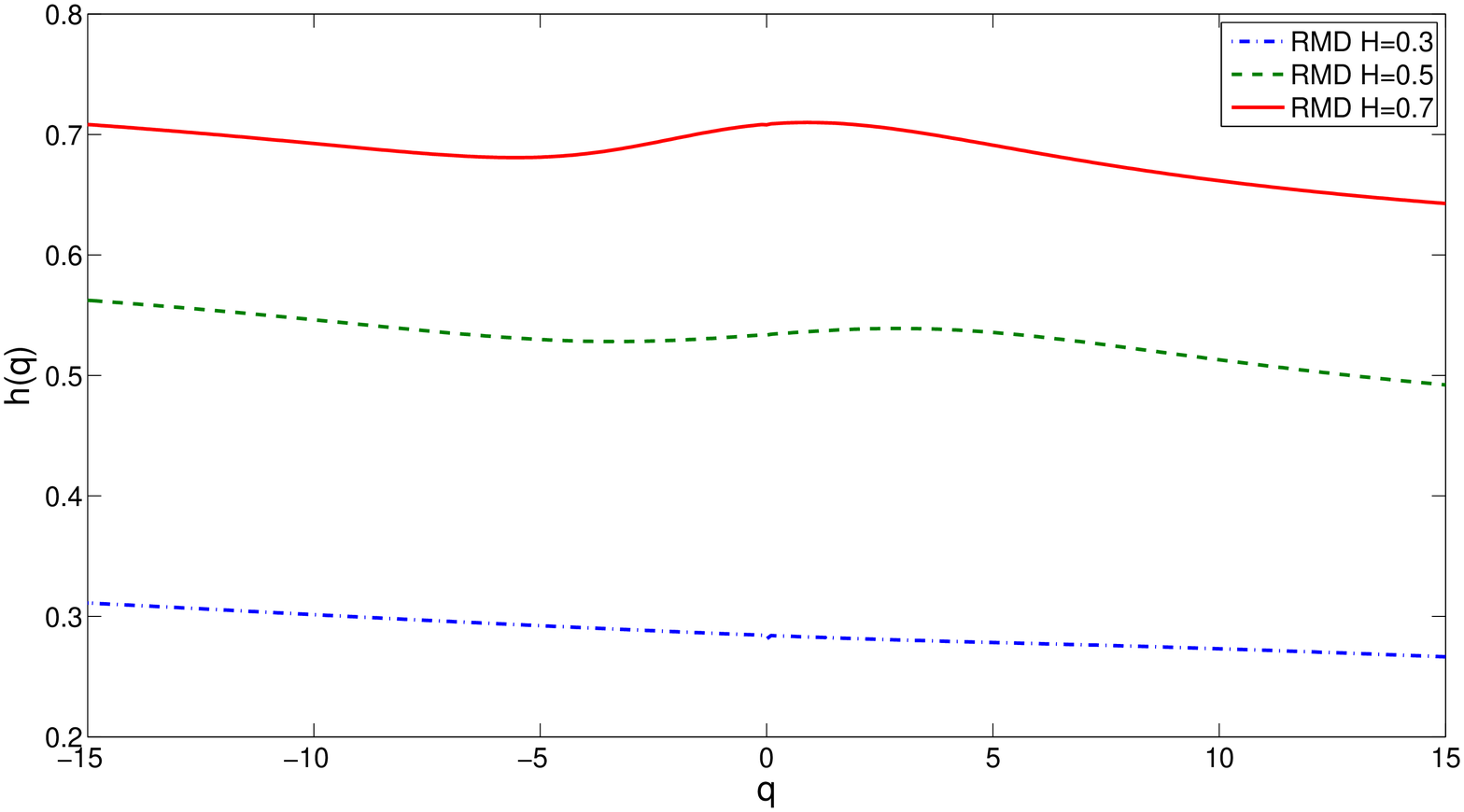}
\caption{The $h(q)$ plot calculated for 3 artificial RMD series simmulated with assumed input values $H$ = 0.3, 0.5 and 0.7 respectively.}
\end{figure}

\begin{figure}[h]
\includegraphics[width=\columnwidth]{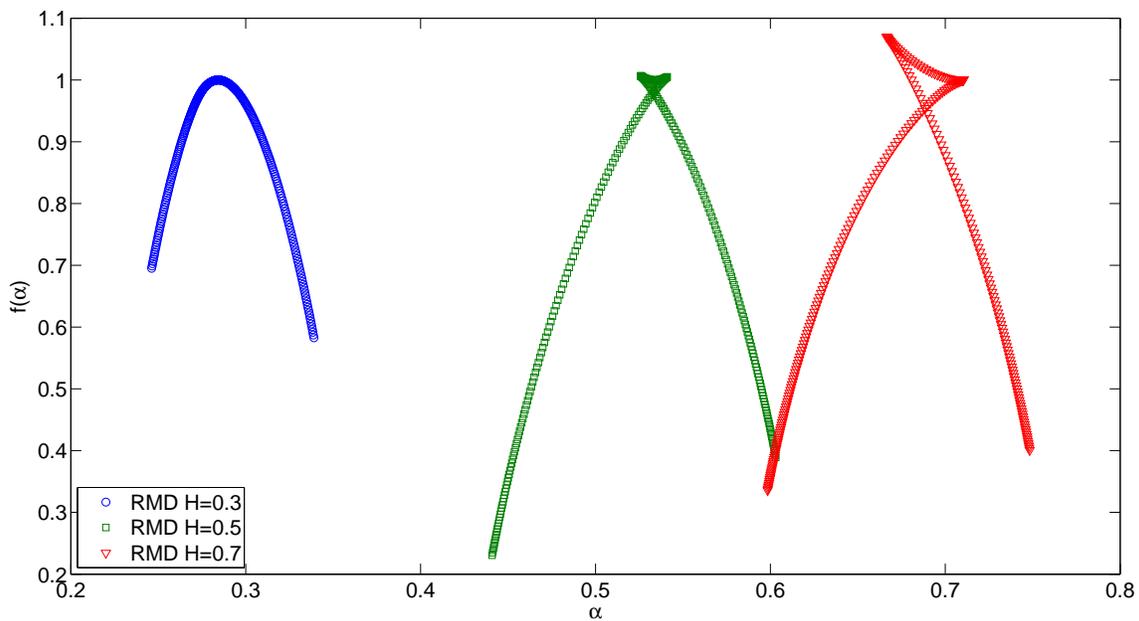}
\caption{Singularity spectra of 3 artificial RMD series with assumed $H$ = 0.3, 0.5 and 0.7 respectively. The peaks of the spectra reflect the pre-set $H$ exponent in terms of the  H\"older exponent $\alpha$.}
\end{figure}

\begin{figure}[h]
\includegraphics[width=\columnwidth]{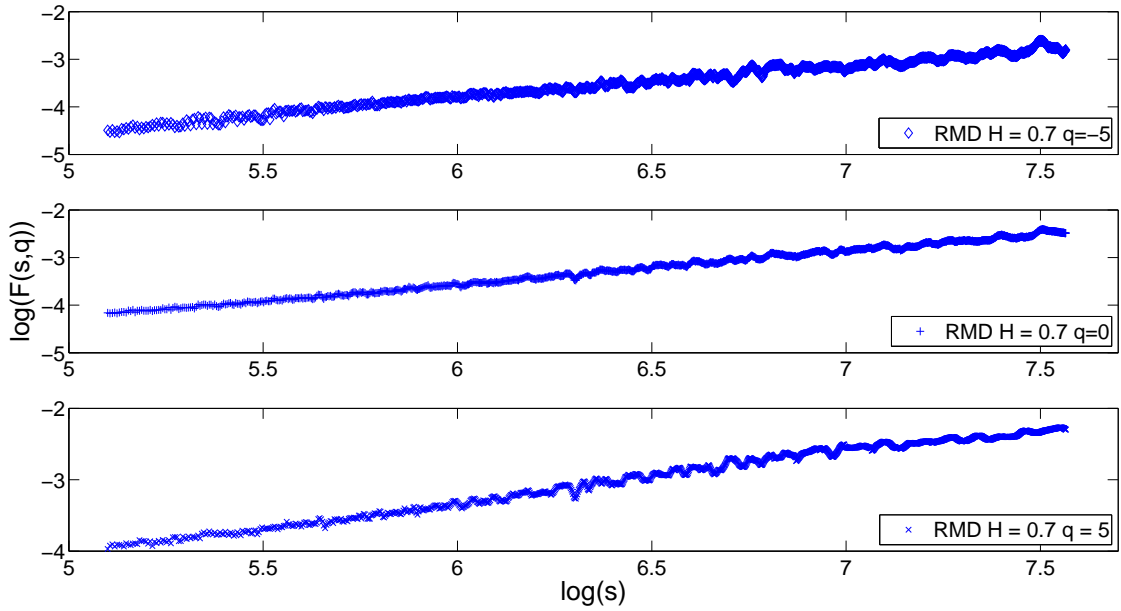}
\caption{An example of $F(s,q) \sim s^{h(q)}$ plot for a series with assumed $H = 0.7$ for $q=-5$ (top), $q=0$ (middle) and $q=5$ (bottom)}
\end{figure}

\begin{figure}[h]
\includegraphics[width=\columnwidth]{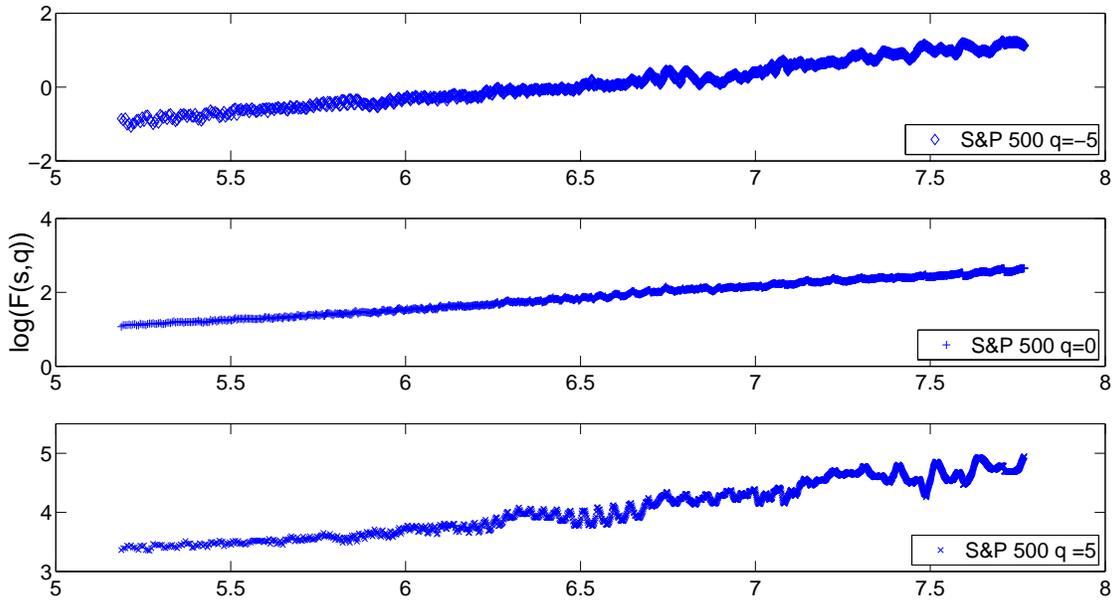}
\caption{Exemplary plots of the $F(s,q) \sim s^{h(q)}$ relation for the entire S\&P 500 series with $q=-5$ (top), $q=0$ (middle) and $q=5$ (bottom).}
\end{figure}

\begin{figure}[h]
\includegraphics[width=\columnwidth]{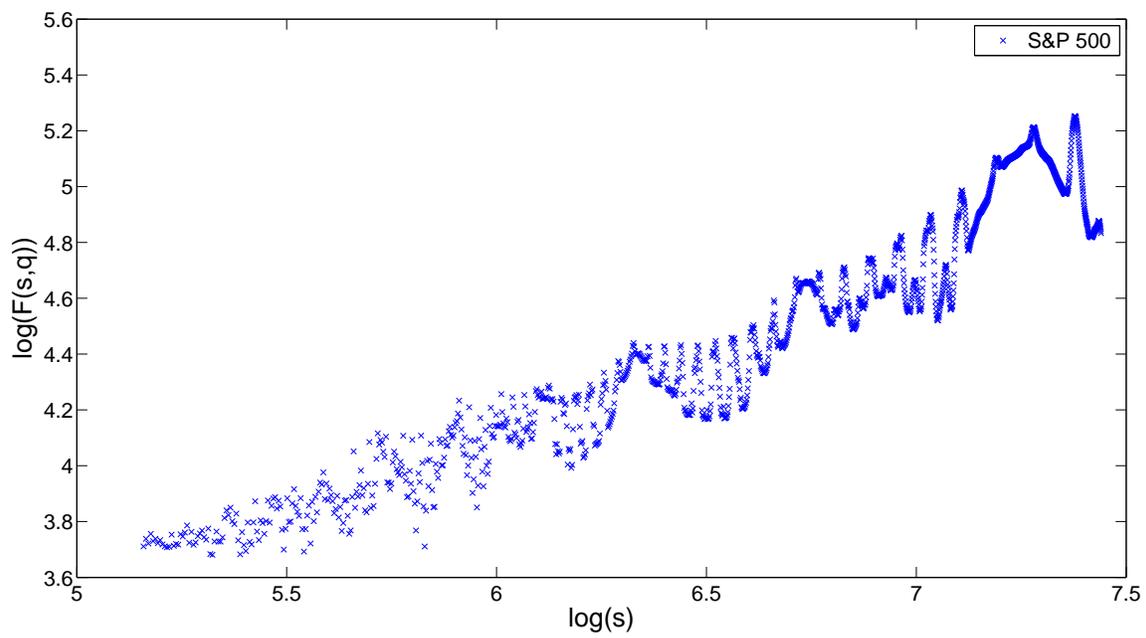}
\caption{Exemplary plot of the $F(s,q) \sim s^{h(q)}$ relation for the entire S\&P 500 series with $q=13$.}
\end{figure}

\begin{figure}[h]
\includegraphics[width=\columnwidth]{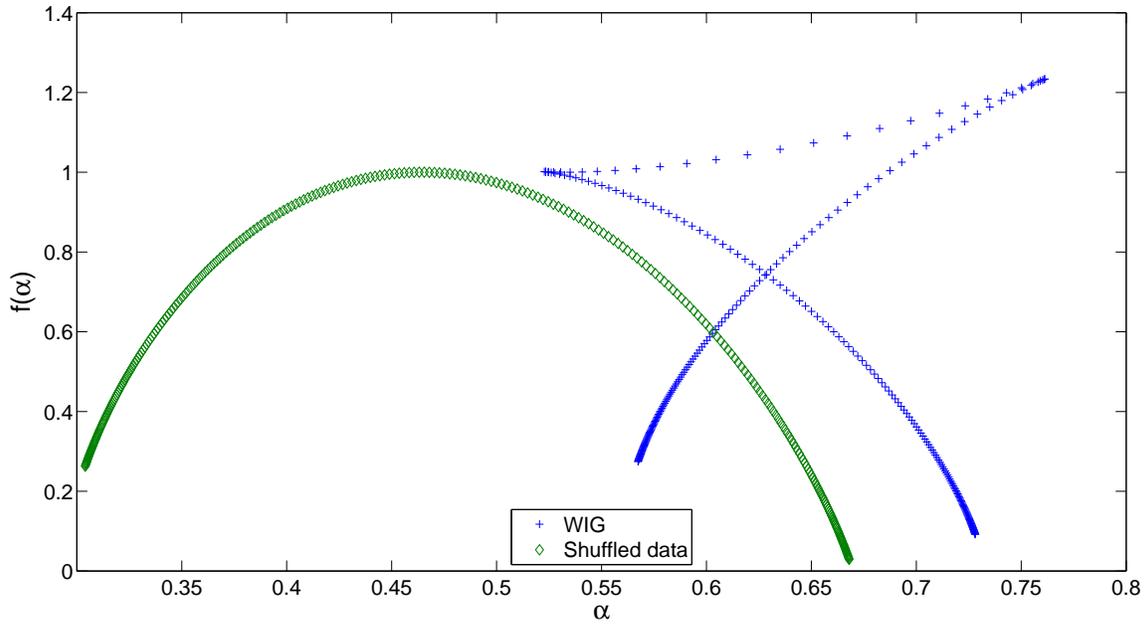}
\includegraphics[width=\columnwidth]{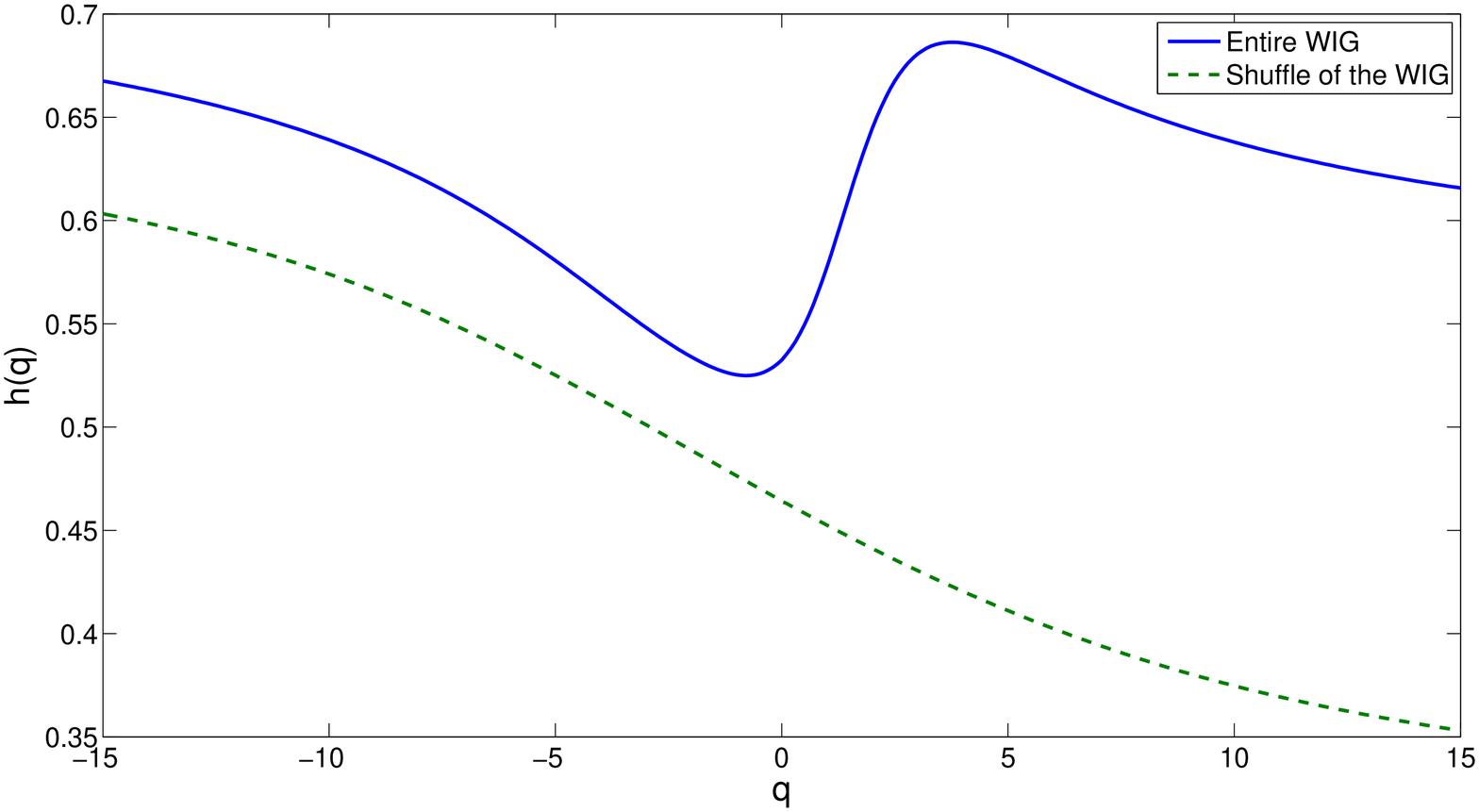}
\caption{The singularity spectra of the entire WIG closing day data and its shuffle. The bottom figure shows the corresponding $h(q)$ dependence.}
\end{figure}

\begin{figure}[h]
\includegraphics[width=\columnwidth]{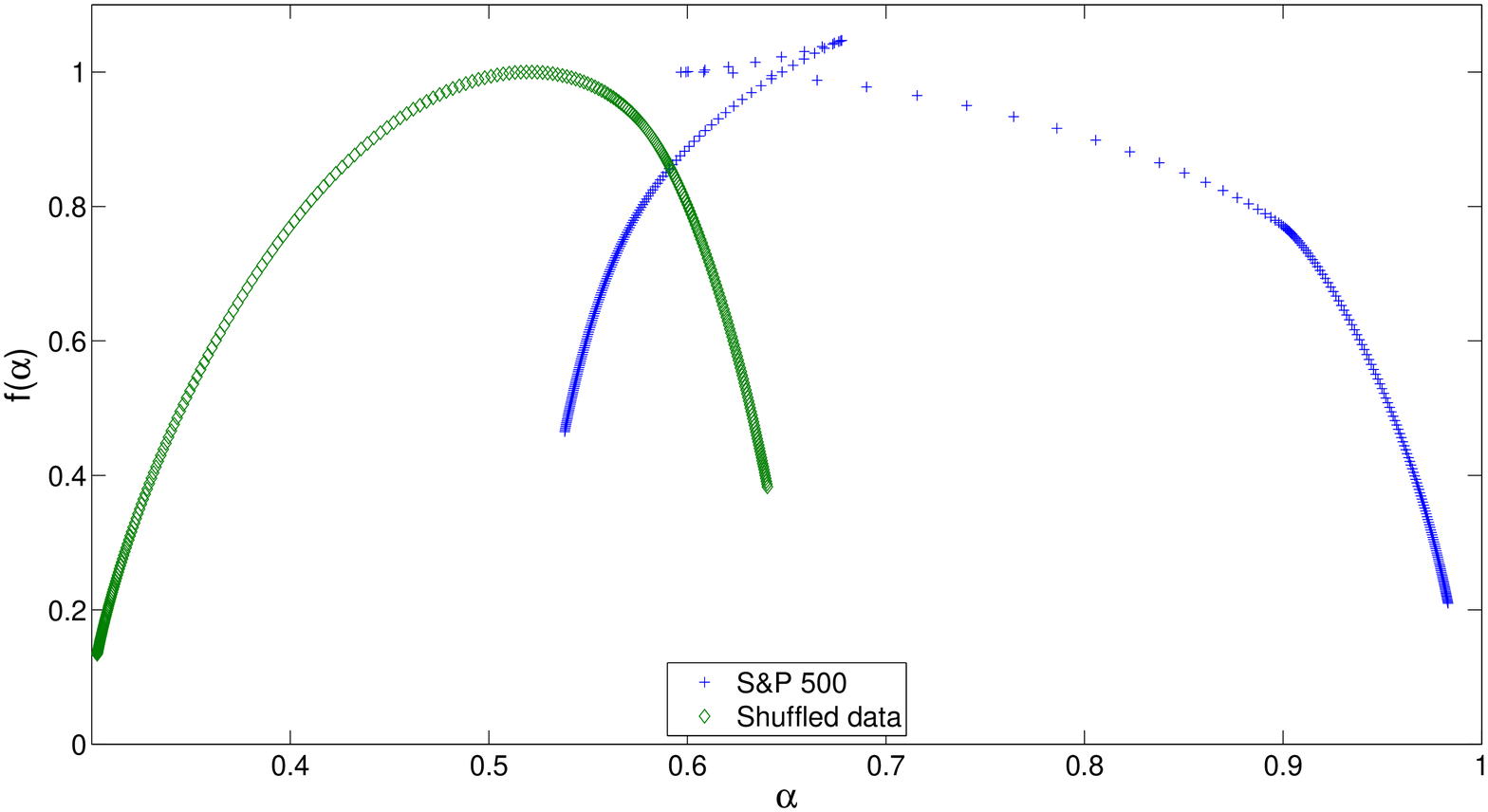}
\includegraphics[width=\columnwidth]{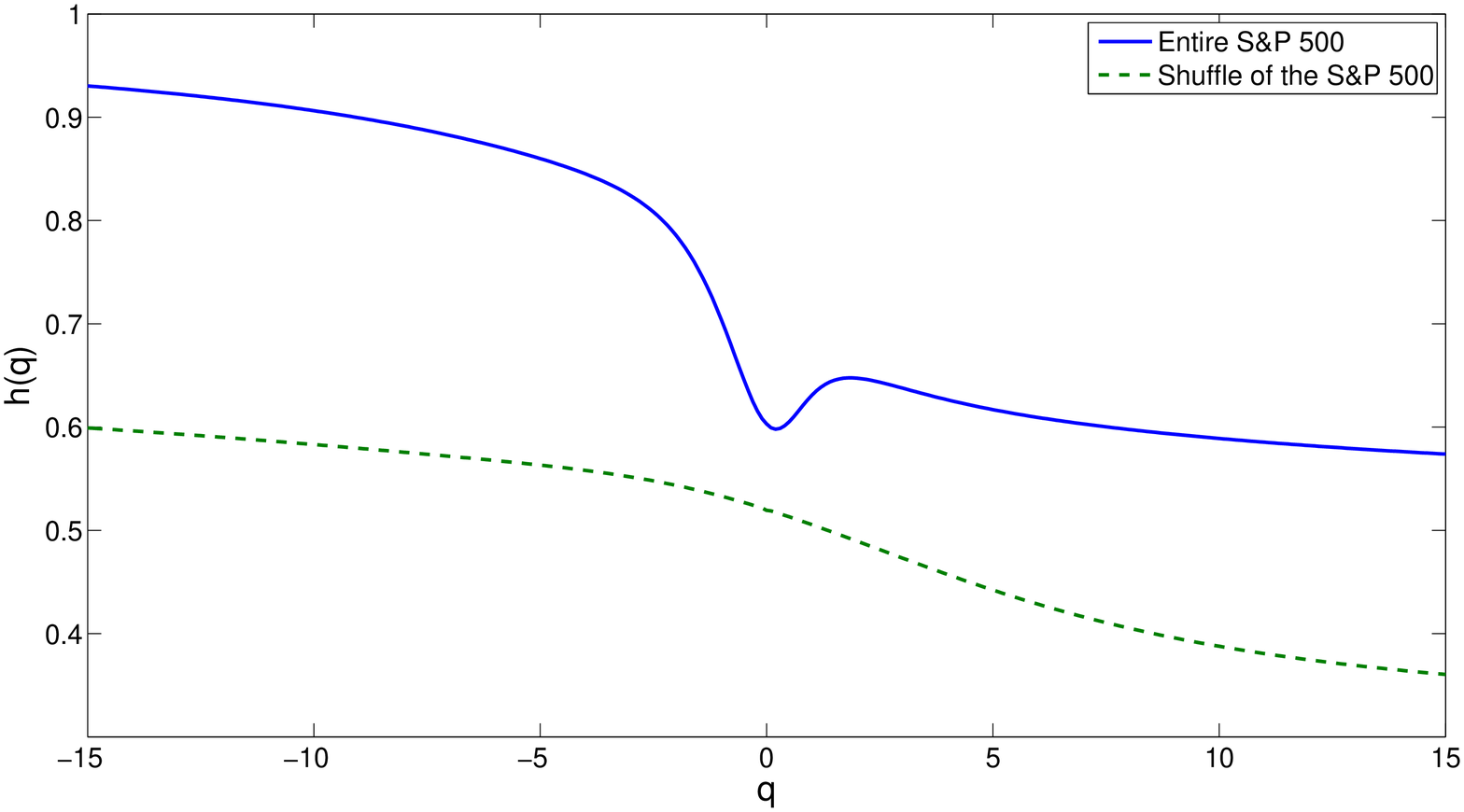}
\caption{The singularity spectra of the entire S\&P500 closing day data and its shuffle. The bottom figure shows the corresponding $h(q)$ plots}
\end{figure}

\begin{figure}[h]
\includegraphics[width=\columnwidth]{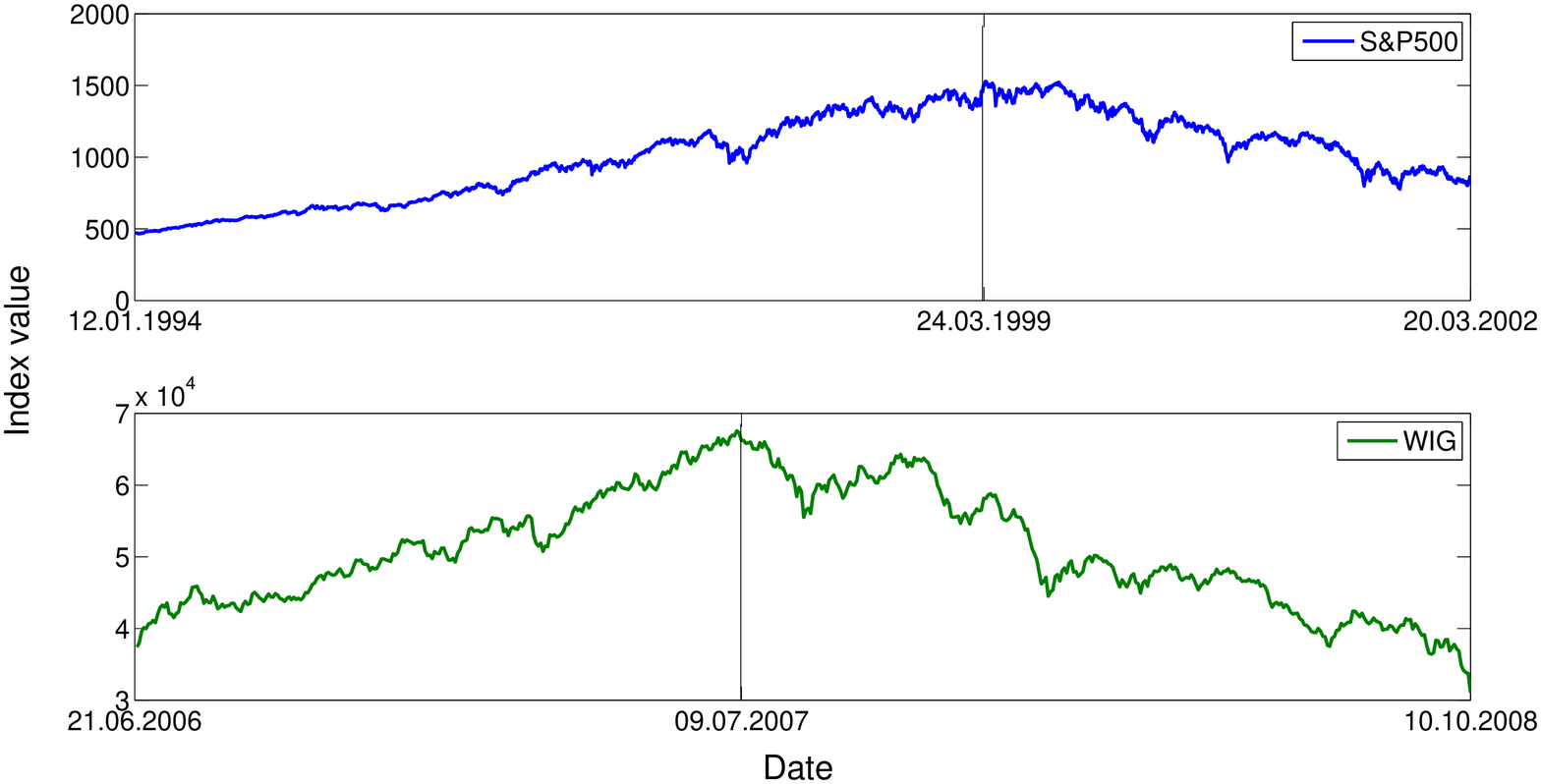}
\caption{Chosen parts of WIG and S\&P500 indices containing the bearish and bullish phases. The moment of the trend change, splitting the series into two distinct phases, is marked as a vertical line.}
\end{figure}

\begin{figure}[h]
\includegraphics[width=\columnwidth]{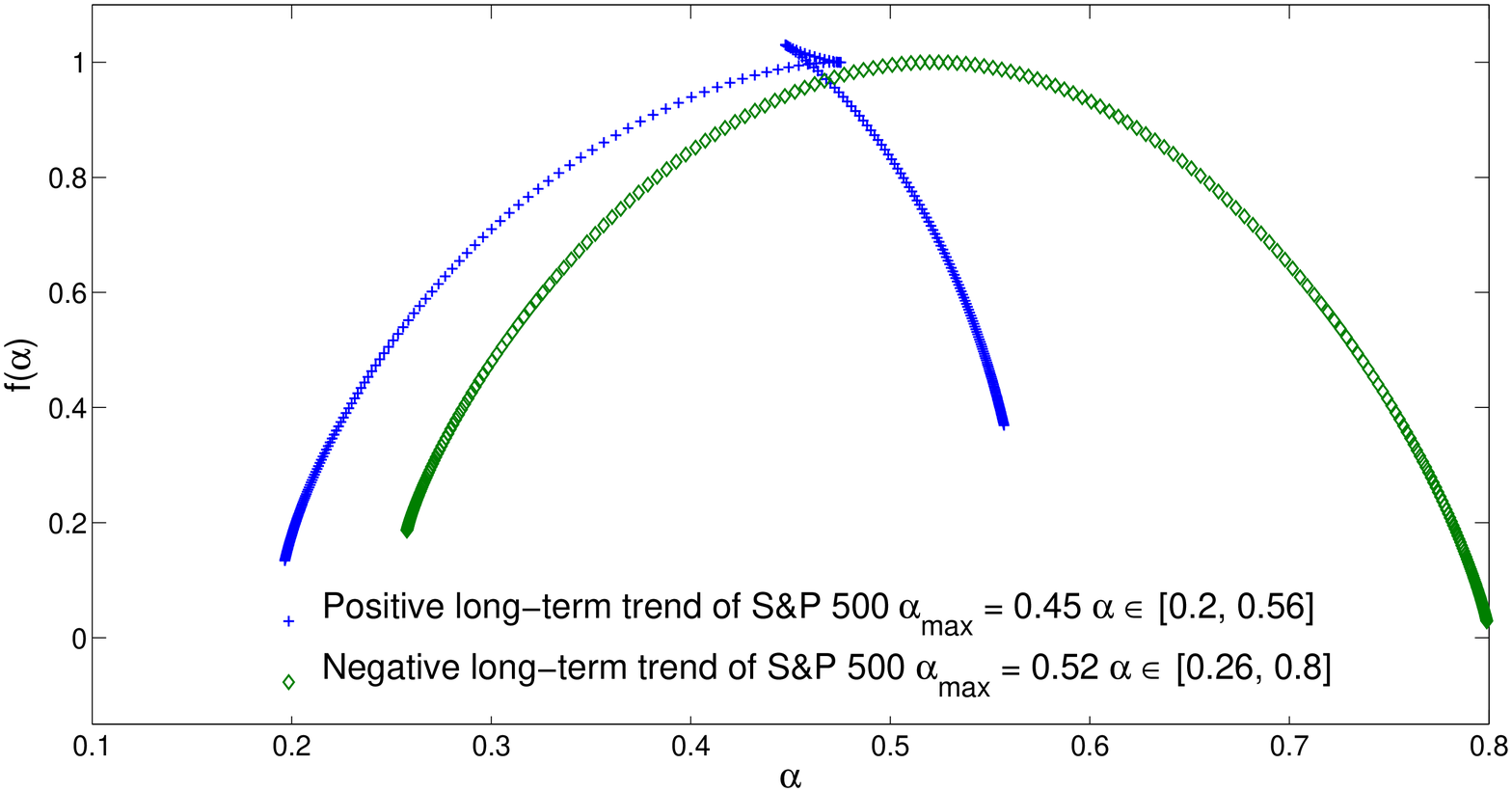}
\includegraphics[width=\columnwidth]{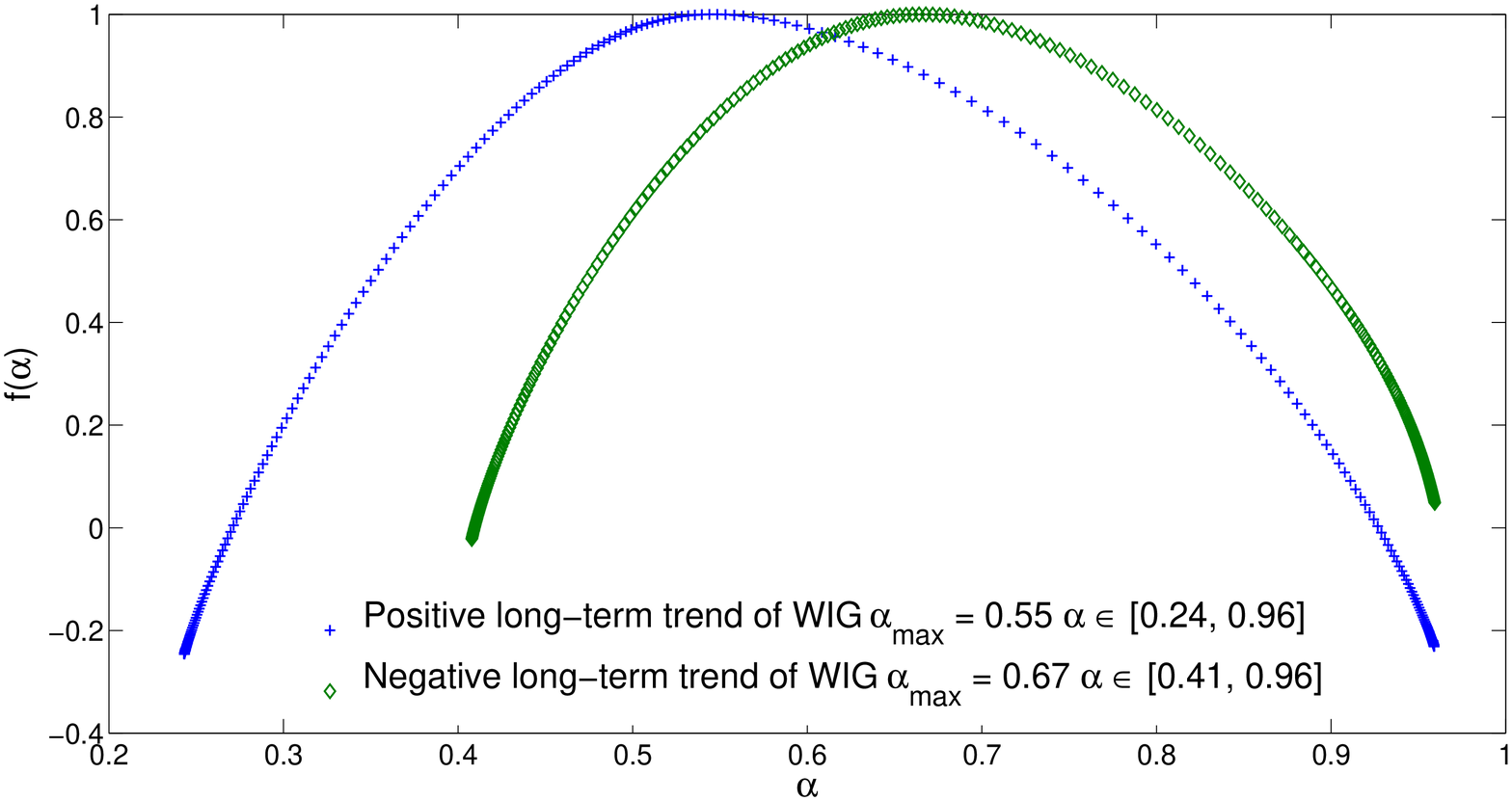}
\caption{Singularity spectra of the bullish and bearish phase for the S\&P500. The bottom figure shows respective plots for WIG index.}
\end{figure}

\begin{figure}[h]
\includegraphics[width=\columnwidth]{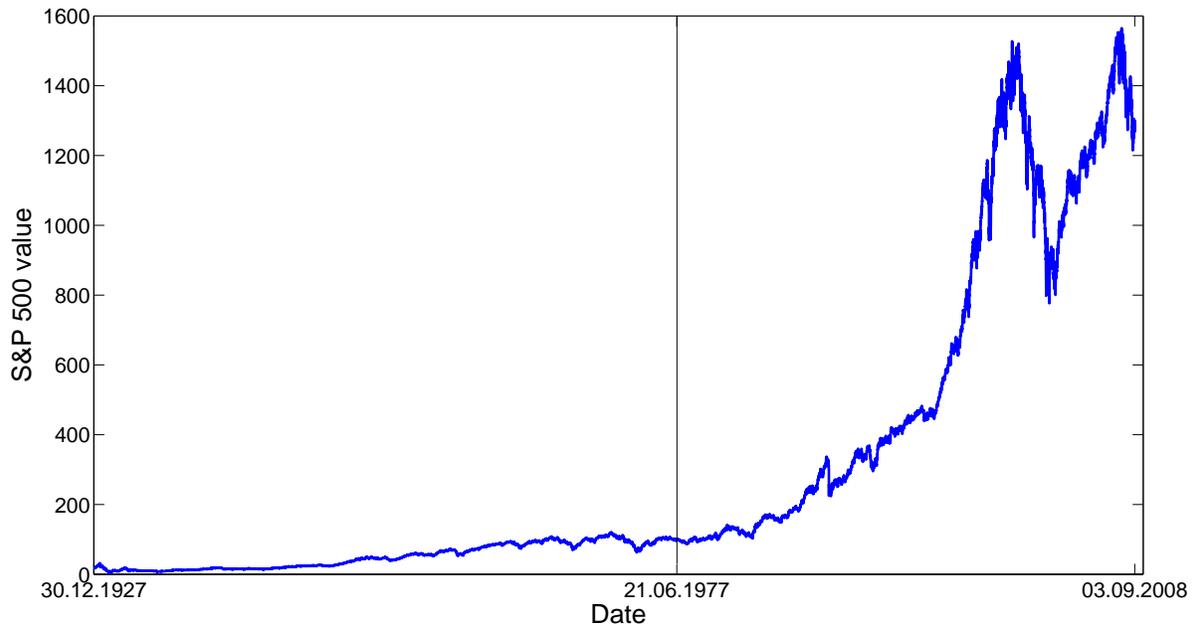}
\caption{The entire S\&P500 time series. The vertical line denotes the border between two regimes. One can see two phases corresponding to low and high oscillations around the trend.}
\end{figure}

\begin{figure}[h]
\includegraphics[width=\columnwidth]{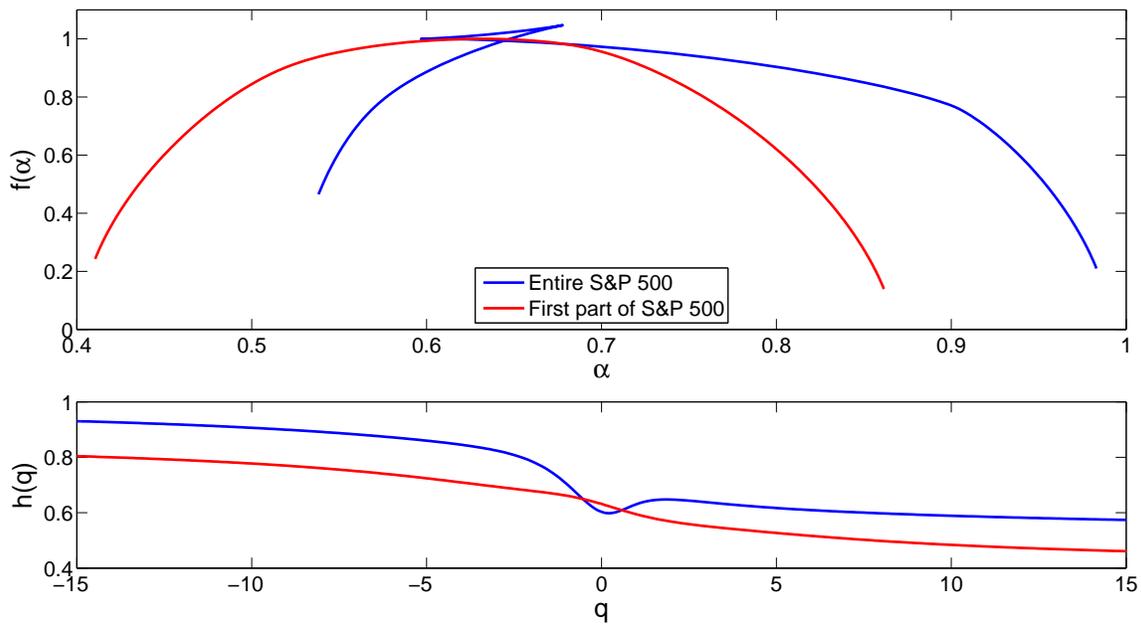}
\caption{Singularity spectra of the first part of the S\&P500 and for the entire signal.}
\end{figure}

\begin{figure}[h]
\includegraphics[width=\columnwidth]{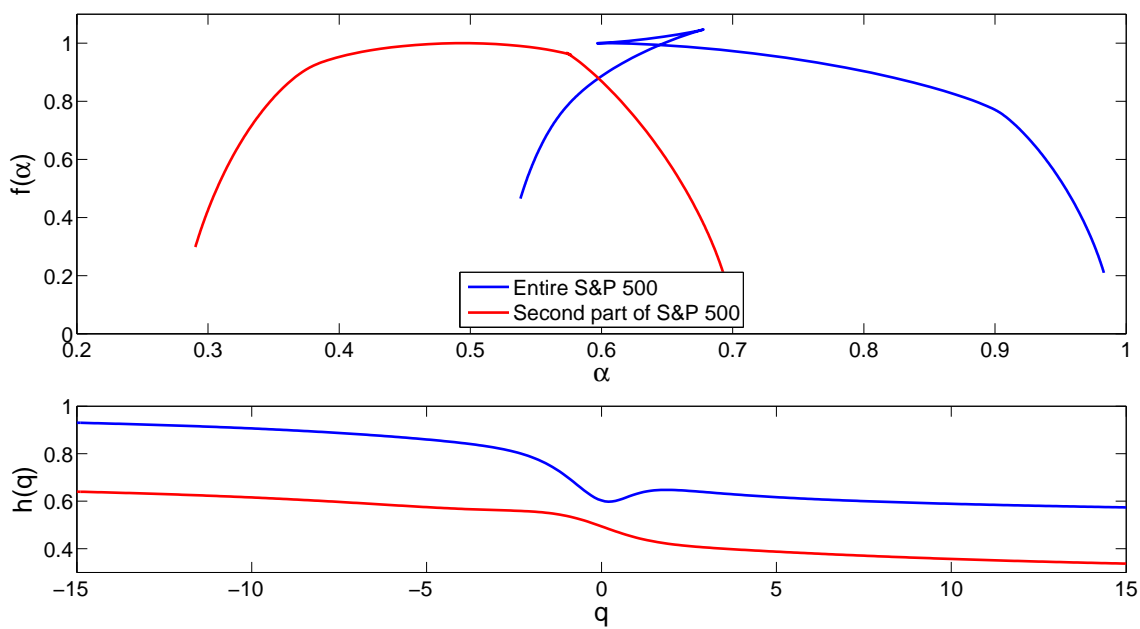}
\caption{Singularity spectra of the second part of the S\&P500 and for the entire signal.}
\end{figure}

\begin{figure}[h]
\includegraphics[width=\columnwidth]{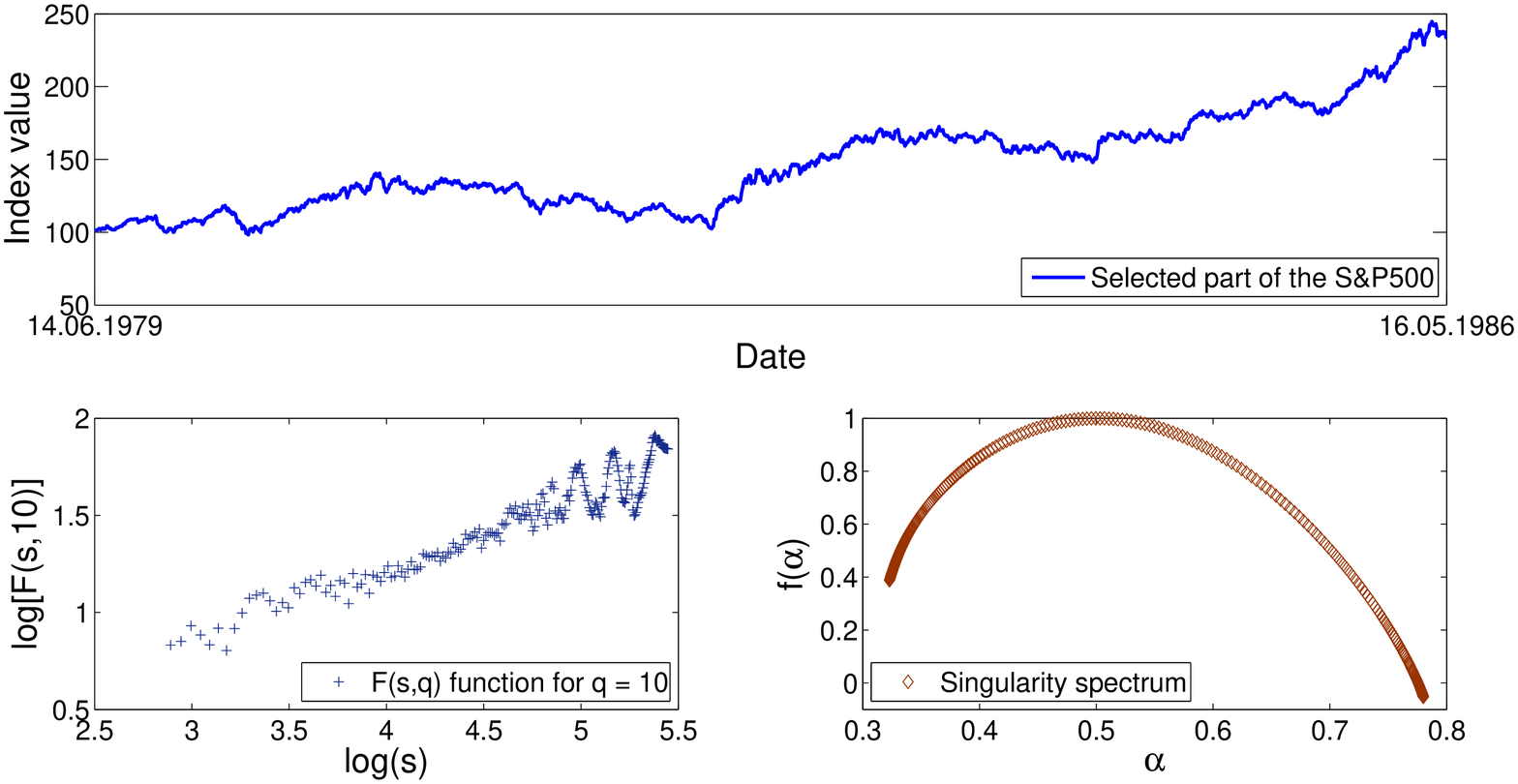}
\caption{The top figure represents an example of ``good'' data without abrupt events. The bottom left plot shows an example of the $F(s,q)$ function scaling for this data. The bottom right figure shows the corresponding singularity spectrum.}
\end{figure}

\begin{figure}[h]
\includegraphics[width=\columnwidth]{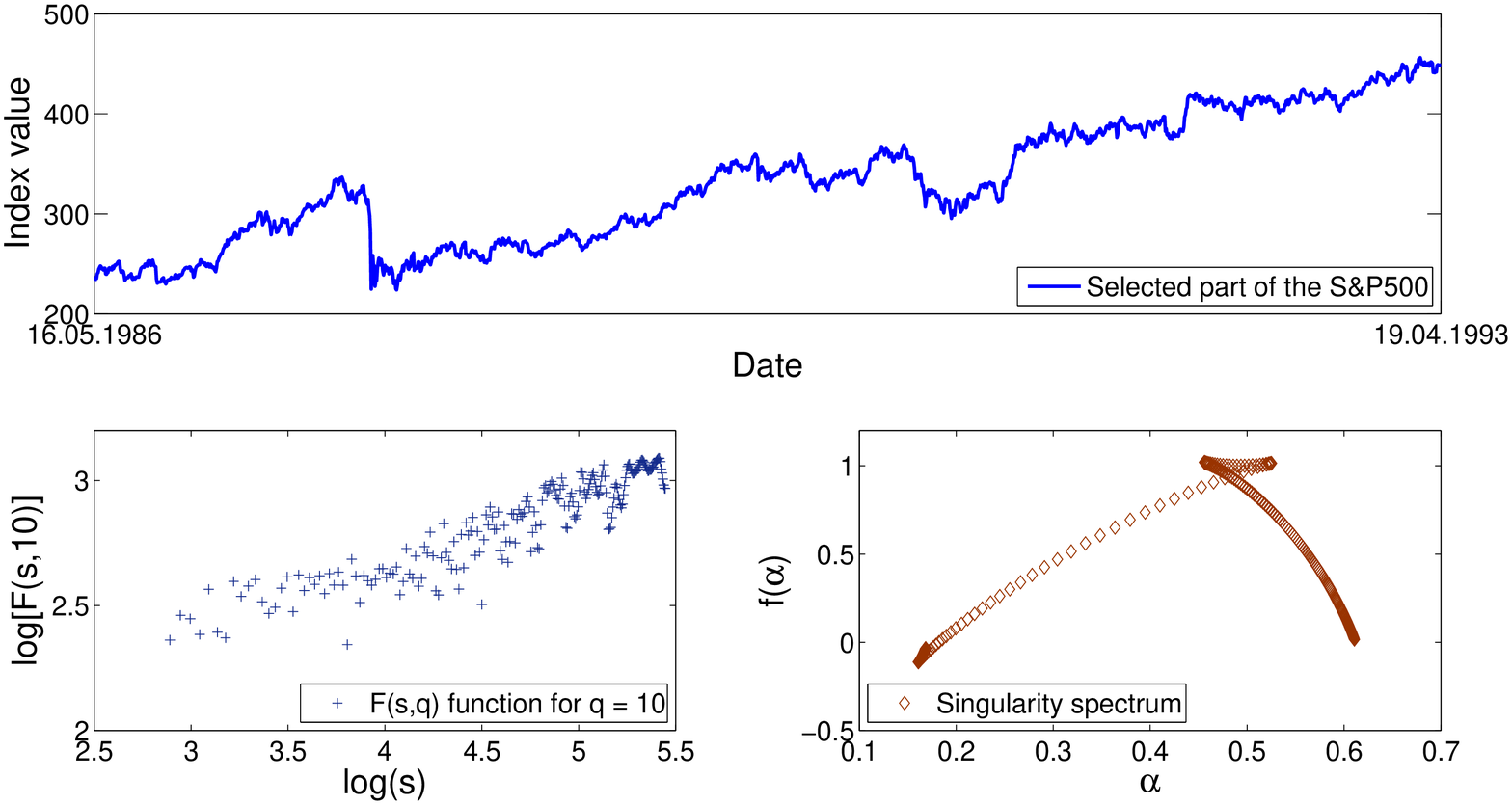}
\caption{The top figure represents an example of ``bad'' data taken immediately after the period shown in Fig.13. On the bottom left an example of $F(s,q)\sim s^{h(q)}$ scaling for this data is presented. It shows much poorer scaling properties than one in Fig.13. The bottom right figure shows the ``twisted'' singularity spectrum, a result of a non-monotonic $h(q)$ dependence.}
\end{figure}

\begin{figure}[h]
\includegraphics[width=\columnwidth]{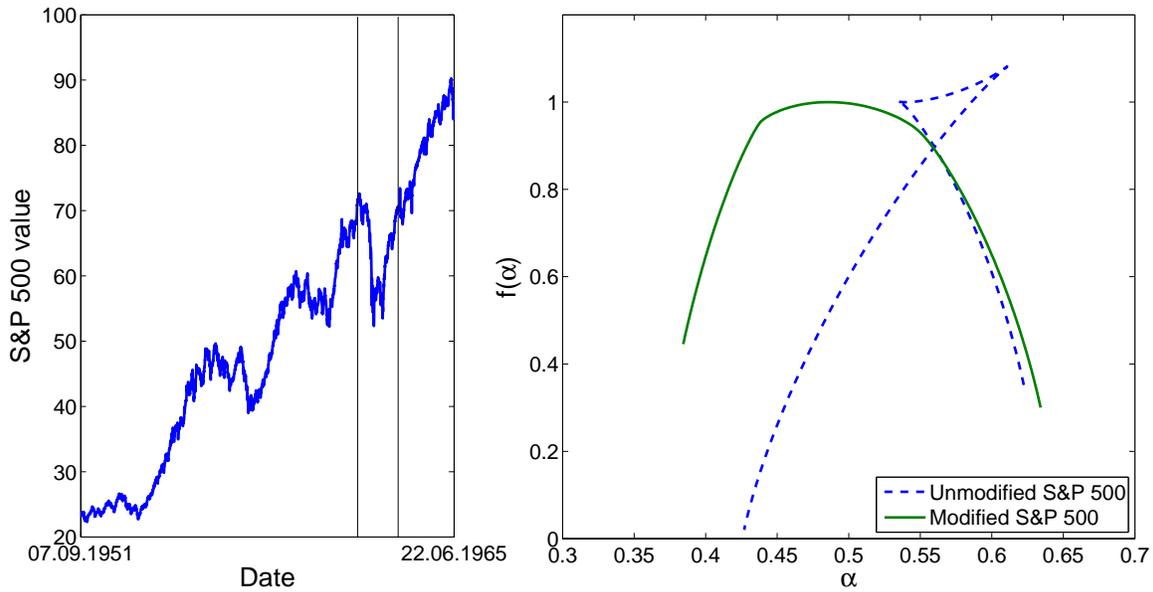}
\caption{The marked region with vertical lines on the left shows the extreme event that was removed from the original S\&P500 signal. The singularity spectra of the original part of the S\&P500 and the modified one is shown on the right.}
\end{figure}

\begin{figure}[h]
\includegraphics[width=\columnwidth]{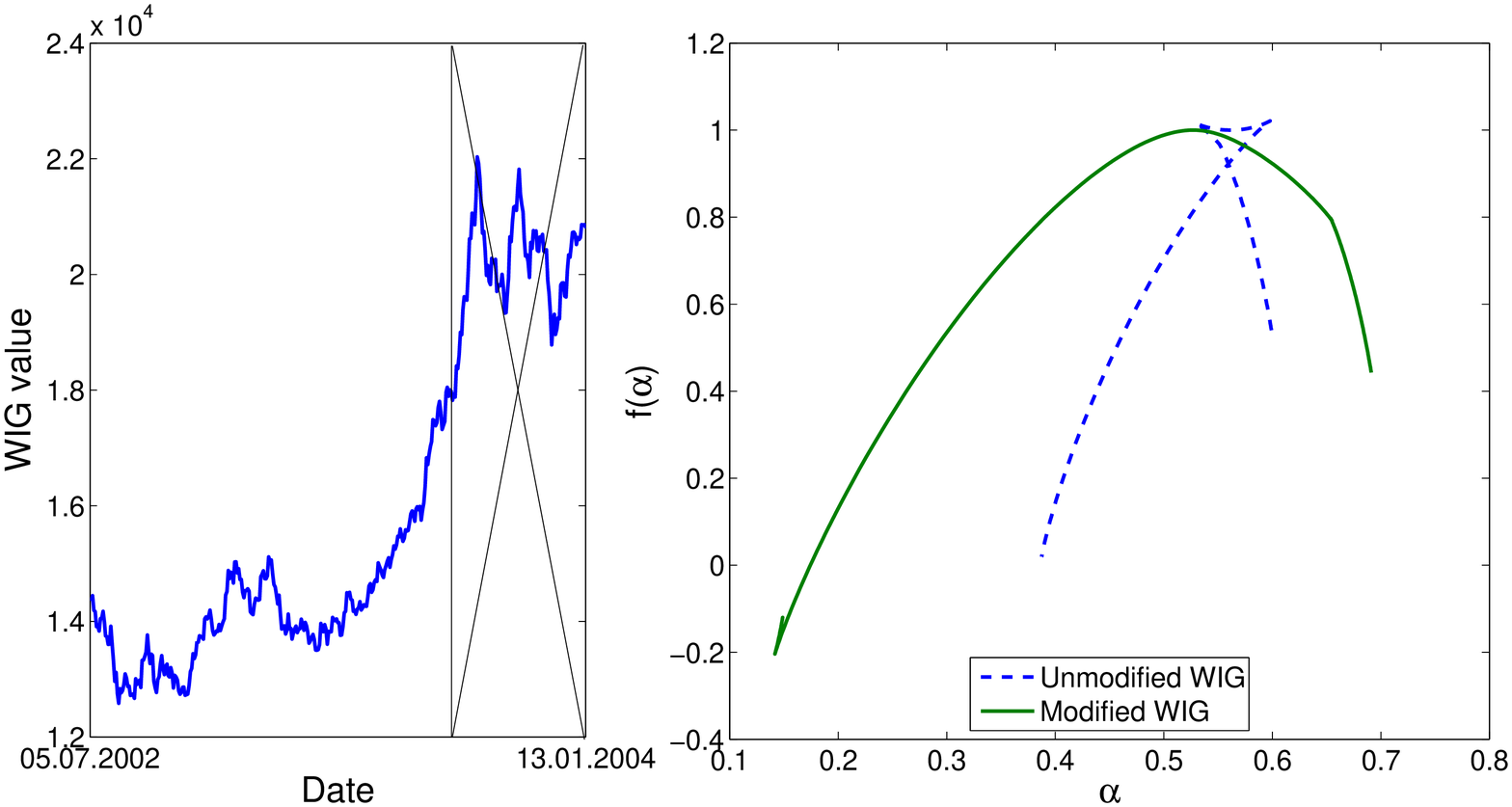}
\caption{The crossed region shows the abrupt regime. On the right, the singularity spectra of the original WIG fragment and the modified spectra after the abrupt regime has been removed.}
\end{figure}

\begin{figure}[h]
\includegraphics[width=\columnwidth]{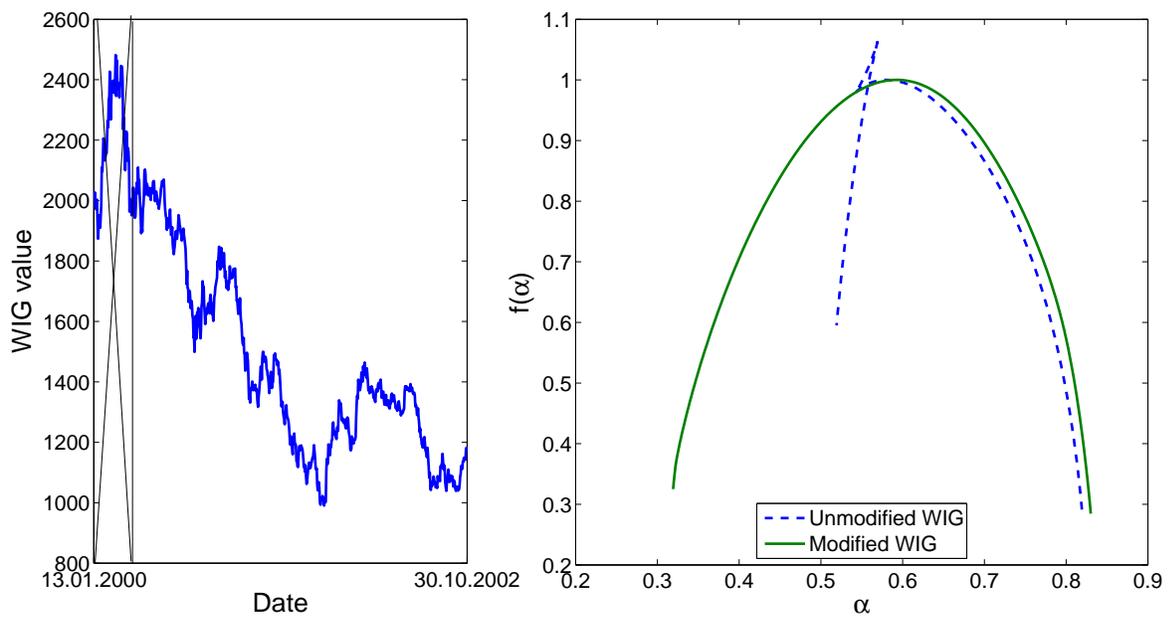}
\caption{Another example of abrupt regime removed from WIG time series (left) and the corresponding change in singularity spectra (right).}
\end{figure}
\end{document}